\documentclass{article}

\usepackage{microtype}
\usepackage{graphicx}
\usepackage{subfigure}
\usepackage{booktabs} 

\usepackage{hyperref}

\usepackage[accepted]{icml2024}

\usepackage{amsmath}
\usepackage{amssymb}
\usepackage{mathtools}
\usepackage{amsthm}

\usepackage[capitalize,noabbrev]{cleveref}

\theoremstyle{plain}

\theoremstyle{definition}

\theoremstyle{remark}

\usepackage[textsize=tiny]{todonotes}

\icmltitlerunning{Multi-Fidelity Bind (MFBind)}

\begin{document}

\twocolumn[
\icmltitle{MFBind: a Multi-Fidelity Approach \\
for Evaluating Drug Compounds in Practical Generative Modeling}


\icmlsetsymbol{equal}{*}

\begin{icmlauthorlist}
\icmlauthor{Peter Eckmann}{ucsd-cs}
\icmlauthor{Dongxia Wu}{ucsd-cs}
\icmlauthor{Germano Heinzelmann}{germano}
\icmlauthor{Michael K Gilson}{ucsd-chem,ucsd-pharm}
\icmlauthor{Rose Yu}{ucsd-cs}
\end{icmlauthorlist}

\icmlaffiliation{ucsd-cs}{Department of Computer Science and Engineering, University of California San Diego, La Jolla, California, United States}
\icmlaffiliation{ucsd-chem}{Department of Chemistry and Biochemistry, University of California San Diego, La Jolla, California, United States}
\icmlaffiliation{ucsd-pharm}{Skaggs School of Pharmacy and Pharmaceutical Sciences, University of California San Diego, La Jolla, California, United States}
\icmlaffiliation{germano}{Departamento de Física, Universidade Federal de Santa Catarina, Florianópolis, Santa Catarina, Brazil}

\icmlcorrespondingauthor{Michael Gilson}{mgilson@health.ucsd.edu}
\icmlcorrespondingauthor{Rose Yu}{roseyu@ucsd.edu}

\icmlkeywords{drug discovery, molecular generation, multi-fidelity learning}

\vskip 0.3in
]



\printAffiliationsAndNotice{} 

\begin{abstract}

Current generative models for drug discovery primarily use molecular docking to evaluate the quality of generated compounds. However, such models are often not useful in practice because even compounds with high docking scores do not consistently show experimental activity. More accurate methods for activity prediction exist, such as molecular dynamics based binding free energy calculations, but they are too computationally expensive to use in a generative model. We propose a multi-fidelity approach, Multi-Fidelity Bind (MFBind), to achieve the optimal trade-off between accuracy and computational cost. MFBind integrates docking and binding free energy simulators to train a multi-fidelity deep surrogate model with active learning. Our deep surrogate model utilizes a pretraining technique and linear prediction heads to efficiently fit small amounts of high-fidelity data. We perform extensive experiments and show that MFBind (1) outperforms other state-of-the-art single and multi-fidelity baselines in surrogate modeling, and (2) boosts the performance of generative models with markedly higher quality compounds.
\end{abstract}

\section{Introduction}
Generative models for \textit{de novo} drug design have gained significant interest in machine learning for their promised ability to quickly generate new compounds. However, generating compounds with real-world activity remains a fundamental challenge \cite{handa2023difficulty, coley2020autonomous}, limiting the widespread adoption of generative models in practical drug discovery \cite{paul2021artificial}. One of the main difficulties is the computational evaluation of compound binding affinity. The generated compounds are often highly novel, so an activity predictor trained with existing experimental data is insufficient due to poor out-of-distribution generalization \cite{chatterjee2023improving, ji2022drugood}. Instead, physics-based methods that model the 3D interaction between compound and target are commonly used.

Due to its speed, molecular docking is the prevalent physics-based method used to evaluate novel compounds by generative models \cite{eckmann2022limo, jeon2020autonomous, lee2023exploring, noh2022path, fu2022reinforced, spiegel2020autogrow4, peng2022pocket2mol, guan20233d, guan2023decompdiff}. However, docking is known to be a relatively poor predictor of activity \cite{pinzi2019molecular, handa2023difficulty, coley2020autonomous, feng2022absolute}, so it would be desirable to apply more accurate binding free energy calculation techniques \cite{pinzi2019molecular, feng2022absolute}. Such techniques, utilizing molecular dynamics simulations, are currently considered the most reliable approach to computational prediction of affinity \cite{moore2023automated, cournia2021free}. However, they have not been used by generative models due to their high computational cost \cite{thomas2023integrating}, with a single compound-protein pair taking hours to days to simulate on a powerful computer \cite{wan2020hit}. Thus, neither docking nor binding free energy techniques alone are sufficient for the practical application of generative models.

Multi-fidelity modeling \cite{fernandez2016review} is an approach to integrate data from simulators with varying accuracy and costs. Multi-fidelity modeling has been successfully applied in scientific areas such as climate modeling \cite{wu2022multi} and materials science \cite{fare2022multi}, but their adoption in drug discovery has been limited. \citet{hernandez2023multi} study their use in peptide design, but they construct a proof-of-concept artificial set of fidelities and costs where even the highest fidelity is not expected to be very accurate.

\begin{figure*}
\begin{center}
\centerline{\includegraphics[width=0.9\textwidth]{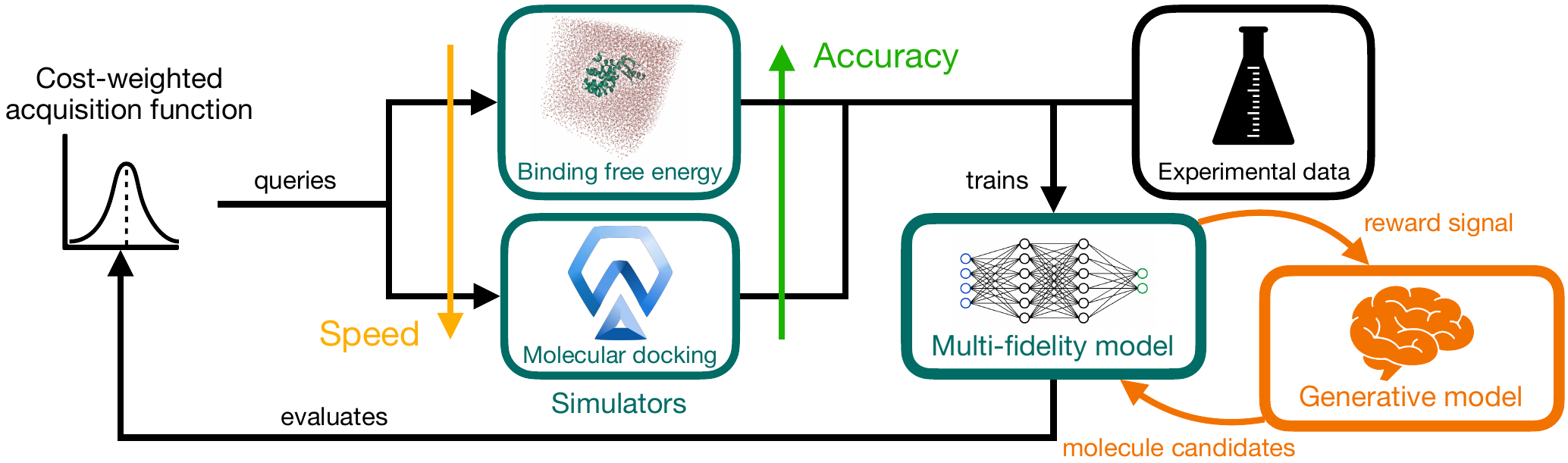}}
\caption{\textbf{Overview of MFBind}. We train a multi-fidelity surrogate model to predict the outputs from all fidelity simulators. Then, we use the model to evaluate the acquisition function, and then pick the next molecule and fidelity level to query the simulators. The result is then added to the training dataset, and the process is repeated. A generative model uses the trained multi-fidelity surrogate model to evaluate its candidate compounds.}
\label{fig1}
\end{center}
\vskip -0.2in
\end{figure*}

In this paper, we address the difficulty of drug compound binding affinity evaluation by proposing a multi-fidelity modeling framework, Multi-Fidelity Bind (\textbf{MFBind}), to achieve the optimal trade-off between accuracy and computational cost. Our framework (Figure \ref{fig1}) consists of a new multi-fidelity environment for binding affinity prediction, and a deep surrogate model that integrates data from each fidelity level to accurately and cheaply mimic the behavior of the binding free energy method. Our model learns a shared compound encoding across all fidelities, and then uses regularized linear heads to output predictions at each fidelity level. We also pretrain the surrogate model on the large quantity of lower fidelity data, and then fine-tune the model on all fidelities, using active learning. More specifically,

\begin{itemize}
    \item we introduce a novel multi-fidelity modeling framework, \textbf{MFBind}, for evaluating the binding affinity of drug compounds in generative modeling that integrates data from existing experimental results, molecular docking, and binding free energy simulators
    \item we propose a deep surrogate model, which utilizes a pretraining technique on the data from the lower set of fidelities, and efficiently train it using a cost-aware multi-fidelity active learning approach.
    \item we perform extensive evaluations of ours and baseline models on two real-world problem settings, multi-fidelity surrogate modeling and generative modeling, showing the practicality of our framework.
\end{itemize}

\section{Related Work}
\subsection{Molecular generative models}
Generative models in drug discovery have gained much interest for their ability to quickly generate compounds with desired properties \cite{paul2021artificial}. Early works \cite{jin2018junction, gomez2018automatic, you2018graph} focus on properties such as the octanol-water partition coefficient (logP) or quantitative estimate of drug-likeness (QED), which are of very limited practical utility \cite{coley2020autonomous, xie2021mars}. More recently, there has been an understanding that the binding affinity to a targeted protein is much more relevant for practical drug discovery \cite{xie2021mars, eckmann2022limo, fu2022reinforced}.

One approach to guide generative models in optimizing compound binding affinity is to use a reward function for compound evaluation. This reward function can be applied to reinforcement learning \cite{jeon2020autonomous, fu2022reinforced}, VAEs \cite{eckmann2022limo, noh2022path}, genetic algorithms \cite{spiegel2020autogrow4, fu2022reinforced}, or diffusion models \cite{lee2023exploring}. All of them use docking software, such as AutoDock \cite{morris2009autodock4}, as the reward function, because it is the only reasonably fast option. However, docking is known to be inaccurate \cite{pinzi2019molecular}, and compounds with high docking scores do not consistently show experimental activity \cite{handa2023difficulty, coley2020autonomous, feng2022absolute}. 

More reliable molecular dynamics-based binding free energy calculations, which are much more accurate than docking \cite{moore2023automated, cournia2021free}, have not yet been applied to \textit{de novo} generative drug design due to their high computational cost \cite{thomas2023integrating}. While \citet{ghanakota2020combining} use binding free energy calculations in combination with a molecular generative model, they focus on the optimization of an existing known lead compound. This allows them to rely on much cheaper relative binding free energy calculations, as opposed to the absolute binding free energy (ABFE) calculations needed for \textit{de novo} design \cite{cournia2017relative}.

Structure-based generative models are trained on 3D structures of protein-ligand pairs, and aim to predict a 3D ligand that fits in a given protein pocket with high binding affinity. Techniques include autoregressive generation \cite{peng2022pocket2mol} and diffusion modeling \cite{guan20233d, guan2023decompdiff}. Despite not needing a reward function like docking during the generation process, the generated compounds are still evaluated with docking as a post-processing step. This means structure-based generative models do not avoid the issue of inaccurate binding affinity prediction.

\subsection{Multi-fidelity modeling}

Multi-fidelity modeling methods aim to fuse multiple data sources of variable accuracy and cost \cite{fernandez2016review}, and are widely used in scientific fields for surrogate modeling and uncertainty quantification \cite{brevault2020overview}. A popular choice of surrogate model is a Gaussian process (GP), which performs well in low data settings and produces well-calibrated uncertainty estimates \cite{brevault2020overview}. One such technique to apply GPs to multi-fidelity modeling is described by \citet{wu2020practical}, where a downsampling kernel is used to output predictions at each fidelity level. While GPs cannot scale to large amounts of data, approaches like KISS-GP \cite{wilson2015kernel} and DKL \cite{wilson2016deep} learn a deep neural kernel to scale GPs. Other surrogate modeling approaches utilize neural processes \cite{wang2020mfpc, wu2022multi} and ordinary differential equations \cite{li2022infinite}.

Multi-fidelity modeling is frequently used in an active learning context, where one uses an estimate of a model's uncertainty to most efficiently acquire more datapoints \cite{ren2021survey}. In the multi-fidelity setting, this means iteratively querying across both the sampling space of molecules and each different fidelity level \cite{li2020deep, hernandez2023multi}. This approach has been applied in climate modeling \cite{wu2022multi}, fluid dynamics \cite{li2020deep, wang2021multi}, and materials science \cite{fare2022multi}. For drug discovery, \citet{hernandez2023multi} seek to design peptides with anti-microbial activity using multi-fidelity active learning, but they construct an artificial hierarchy of fidelities and associated costs by training the same machine learning model on different data subsets. This means even the highest fidelity is likely to not be very accurate, as machine learning predictors are typically not as accurate as physics-based methods \cite{moore2023automated, cournia2021free, chatterjee2023improving} and suffer from out-of-distribution generalization issues \cite{ji2022drugood}.
\section{MFBind}
We introduce \textbf{MFBind}, a novel multi-fidelity modeling framework for evaluating compound binding affinities. Our framework consists of a multi-fidelity environment, a deep surrogate model, and an active learning algorithm.

\subsection{Multi-fidelity binding affinity environment}
A multi-fidelity environment consists of a set of simulators $\{f_1, \cdots, f_K\}$, each of which output an increasingly accurate estimate of the value of interest. Define $c>0$ as the computational cost for a given simulator, such that $c_1 < c_2 < \cdots < c_K$. The goal of multi-fidelity modeling is to learn a surrogate model $\hat{f}_K$ that can accurately approximate $f_K$ using a limited amount of high-fidelity data by incorporating data from multiple simulators.

We introduce a new multi-fidelity environment for binding affinity which uses three simulators, each of which takes a molecule as input and outputs an estimate of its binding affinity to a targeted protein with increasing accuracy:
\begin{enumerate}
    \item \textbf{AutoDock4} ($f_1$; $c_1=30s$) \cite{morris2009autodock4}. Uses geometric and charge information from the protein and compound to estimate the binding energy. It outputs a total binding energy prediction in kcal/mol, as well as a set of 15 other outputs, such as energy components for each type of interaction and the number of protein-compound hydrogen bonds (see Appendix \ref{appendix-env} for a full list), some of which are computed in a post-processing step by BINANA \cite{young2022binana}.
    \item \textbf{Experimental data} ($f_2$; $c_2=N/A$) \cite{liu2007bindingdb}. Binding values from laboratory experimental studies, obtained from BindingDB. Because it is infeasible to ``query'' a laboratory for new compounds, we restrict this simulator to only evaluate compounds with known activity values.
    \item \textbf{Absolute binding free energy} (ABFE) ($f_3$; $c_3=37,521s=10.4 hrs$) \cite{heinzelmann2021automation}. A binding free energy method applicable to \textit{de novo} discovery that uses molecular dynamics simulations to accurately predict the binding energy in kcal/mol.
\end{enumerate}
Note that AutoDock4 produces a total of 16 different outputs related to the protein-compound interaction, each of which can be modeled and may aid in the prediction of ABFE scores. The other two fidelities only output a single value. See Appendix \ref{appendix-env} for more details about the environment.

AutoDock4 and ABFE can calculate activities for any compound, whereas the experimental data is only present for a limited set of compounds. While our goal is ultimately to discover compounds with strong experimental binding, we do not make experimental data the highest fidelity simulator because it cannot be queried for arbitrary molecular inputs. Instead, we treat ABFE as the highest fidelity simulator, and use the limited experimental data as a way to improve the surrogate modeling of ABFE scores.

\begin{figure}
\begin{center}
\centerline{\includegraphics[width=0.92\columnwidth]{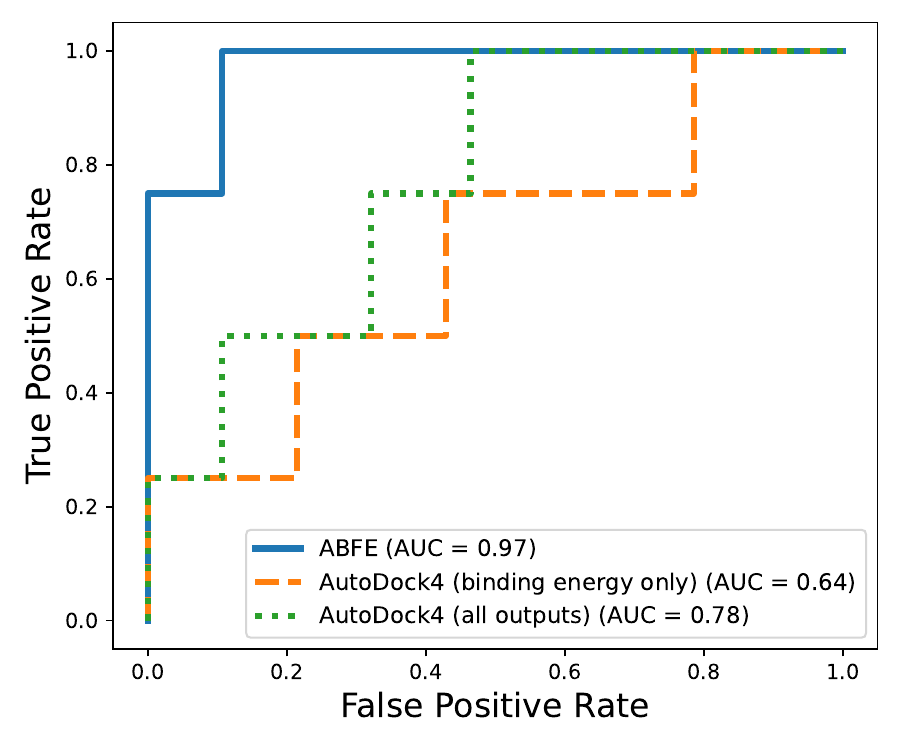}}
\vskip -0.1in
\caption{\textbf{ROC curve of each simulator on the BRD4(2) test set}. Two curves are shown for AutoDock4, one that uses the total binding energy output only, and one that uses a linear surrogate that takes all 16 outputs from AutoDock4 and outputs a prediction of the ABFE score.}
\vskip -0.2in
\label{small-fid-corrs-fig}
\end{center}
\end{figure}

To prove that the higher cost simulators are more accurate, Figure \ref{small-fid-corrs-fig} shows the classification performance of the AutoDock4 and ABFE simulators on the test set for the BRD4(2) target (see Sec. \ref{sec-experiment}). The test set consists of half experimentally confirmed active compounds and half presumed inactives. As expected, the ABFE simulator is the most accurate, while AutoDock4 is still moderately predictive. Additionally, a linear surrogate that uses all 16 AutoDock4 outputs outperforms using only AutoDock4's total binding energy prediction. See Appendix \ref{appendix-result-environment} for more information about these results and data from more targets.

\subsection{Multi-fidelity deep surrogate model}
Many previous multi-fidelity surrogate models are ill-suited to our proposed environment because the simulators at each fidelity have variable output dimensions, and collecting large amounts ($>100$ samples) of high-fidelity data is very costly. Specifically, GP-based methods can only fit a single scalar output at each fidelity level \cite{wu2020practical} and/or do not support a variable number of output dimensions at each fidelity \cite{wang2021multi}. Deep learning models \cite{li2020deep, wu2023disentangled} also struggle to generalize from a limited amount of high-fidelity training data, likely due to over-parameterization of individual layers.


We propose a new deep surrogate model to address these limitations. Our model consists of a neural encoder, shared across all fidelities, to generate a representation of the input molecule, and linear fidelity-specific prediction heads that transform that representation into a prediction of the fidelity output. The prediction heads are able to output varying dimensionalities for each fidelity. 


Mathematically, let $h$ be a feedforward neural network that encodes a molecule $x$, represented as a 2048-dimensional Morgan fingerprint, into an $n$-dimensional real-valued vector. Then, define the fidelity-specific heads as a vector of parameters $\mathbf{w}_1 \in \mathbb{R}^{n \times 16}$ and $\mathbf{w}_2, \mathbf{w}_3 \in \mathbb{R}^{n \times 1}$, and a set of biases, $b_1 \in \mathbb{R}^{16}, b_2, b_3 \in \mathbb{R}$, one for each fidelity level. Note that the weights and biases for the first fidelity level, AutoDock4, are 16-dimensional so that we can model all outputs from that simulator, while the other two fidelities only have scalar output. See Appendix \ref{appendix-mfbind} for more details and a diagram of our surrogate model.

The surrogate model is trained on a dataset $\mathcal{D}$ comprising tuples of the form $(x, k, f_k(x))$, where $x$ is a molecule and $k$ is a fidelity level. The model is trained to minimize the following loss function:

\begin{equation}
\begin{split}
\frac{1}{|\mathcal{D}|} \sum_{x, k, f_k(x) \in \mathcal{D}} &\lambda_k \sum_{i=1}^{|f_k(x)|} \left( ( h(x) \mathbf{w}_k^{(i)} + b_k^{(i)} ) - f_k(x)^{(i)} \right)^2 \\
&+ \lambda_{reg} \sum_{k=1}^3 ||\mathbf{w}_k||_2^2 
\end{split}
\end{equation}

where $f_k(x)^{(i)}$ is the $i$th output of the simulator at fidelity level $k$ (only relevant for AutoDock4, which has 16 outputs), $||_2^2$ denotes L2 regularization, $\lambda_k$ is a weighting parameter for fidelity $k$, and $\lambda_{reg}$ is the regularization strength. The regularization of the linear heads ensures that the model does not overfit on a small amount of high-fidelity data.

We employ a pretraining strategy to improve model performance and reduce overfitting issues, where we first train the model on only the two lowest fidelities (without ABFE), and then finetune on all fidelities, including ABFE. The number of epochs for both phases is determined via hyperparameter tuning. This approach is inspired by the pretraining method for multi-task learning \cite{kaplun2023subtuning}. Intuitively, the pretraining phase helps the model learn an effective encoder on the two fidelities with a large amount of data, without overfitting the encoder on the small number of ABFE points. Then, after this ``warm start'' on the encoder, the finetuning phase allows the model to learn features specific for ABFE prediction.

We use Monte-Carlo (MC) dropout \cite{gal2016dropout} to estimate model uncertainty for active learning (explained in the subsequent section). For AutoDock4, which has a 16-dimensional output, we normalize each of the 16 elements to have a mean of zero and unit variance (across the training dataset), and then average the variance across all elements. We perform this normalization step so that one element with a greater magnitude does not dominate the uncertainty estimation.

\begin{algorithm}[t!]
\caption{Multi-fidelity active learning}
\label{al-algorithm}
\begin{algorithmic}
\REQUIRE a multi-fidelity surrogate model $g$, a pre-populated multi-fidelity training dataset $\mathcal{D}$, a set of candidate points $\mathcal{S}$, the cumulative active learning cost $C \gets 0$, a set of costs $c_1, c_2, c_3$, and the computational budget $B$
\WHILE{$C < B$}
\STATE $g \gets$ TrainSurrogateModel($\mathcal{D}$)
\STATE $maxX, maxK, maxVar \gets 0, 0, 0$
\FOR{$x$ in $\mathcal{S}$}
\FOR{k in 1...3}
\STATE $var \gets a(x, k)$ (Sec. \ref{eq-acq})
\IF{$var > maxVar$}
\STATE $maxX, maxK, maxVar \gets x, k, var$
\ENDIF
\ENDFOR
\ENDFOR
\STATE $\mathcal{D} \gets \mathcal{D} \cup \{ (maxX, maxK, f_{maxK}(maxX)) \}$
\STATE $C \gets C + c_{maxK}$
\ENDWHILE
\end{algorithmic}
\end{algorithm}

\subsection{Active learning to train multi-fidelity surrogate}
Learning a multi-fidelity surrogate model on our proposed environment requires significant computational resources, especially to gather data from ABFE. Instead of passively collecting training data, we propose an active learning approach to efficiently query the simulators.

Our active learning algorithm involves iterative querying of the simulators at the points where the model is most uncertain, weighted by the computational cost. The model is first trained on a limited prepopulated dataset $\mathcal{D}$, and then evaluates its uncertainty $a(x,k)$ on each of the candidate compounds $x$ at each fidelity level $k$. After the compound and associated fidelity level, i.e. simulator, with the highest uncertainty is selected, that compound is run through the simulator and the new activity data is added to the training dataset $\mathcal{D}$. The model is updated with the new data, and the process is repeated until a computational budget $B$ is reached. Algorithm \ref{al-algorithm} describes such a procedure.


To quantify model uncertainty we define an acquisition function $a(x, k)$ that outputs the expected utility of querying the simulator at fidelity $k$ for a compound $x$. In this paper, we use the cost-weighted maximum variance,
$
    \label{eq-acq}
    a(x, k) = \frac{1}{c_k} \sigma^2(x, k)
$.
Here, $\sigma^2(x, k)$ is the model uncertainty (variance) at point $x$ and fidelity $k$. We also tried an entropy-based acquisition function, but found empirically that the cost term dominated the entropy term, and thus the model always chose to query the lowest cost simulator.

Since querying ABFE and docking for a single compound is already parallelizable across multiple GPUs/cores \cite{heinzelmann2021automation}, we do not consider batch active learning algorithms \cite{kirsch2019batchbald}. We chose to retrain the surrogate model after every query because the computational cost of surrogate training is much lower than even the fastest simulator.

\section{Experimental results}

\label{sec-experiment}

To evaluate the utility of our MFBind framework, we consider the following two experimental settings. First, we evaluate the predictive performance of our multi-fidelity surrogate model trained with active learning compared to baseline multi-fidelity techniques. Second, we integrate our deep surrogate model with existing generative models, and compare the generated compounds to those generated by the traditional single-fidelity methods. See Appendix \ref{appendix-experimental-details} for experimental details.

We conduct experiments on two targeted proteins: BRD4(2) (PDB 5UF0) and c-MET (PDB 5EOB). BRD4(2), the second binding domain of bromodomain-containing protein 4, is a protein of interest for cancer treatment \cite{french2008molecular}. c-MET is a receptor tyrosine kinase that also shows promise for treating cancer \cite{zhang2018function}. ABFE has previously been well-validated for both of these targets, where it shows strong correlation with experimental results \cite{heinzelmann2021automation, huggins2022comparing}.

\subsection{Multi-fidelity surrogate modeling}

\subsubsection{Setup}
We first evaluate the performance of our multi-fidelity surrogate model trained with active learning, using Algorithm \ref{al-algorithm} and a predefined candidate set of compounds, and compare the results with those of baseline methods.

The training dataset for this test is initialized with data across all fidelity levels, and then is supplemented with active learning data as the model queries the simulators. The initial dataset for BRD4(2) includes the AutoDock4 output for 100,000 compounds randomly sampled from the ZINC250k dataset \cite{irwin2012zinc}, 91 experimental activity values for the BRD4(2) target obtained from BindingDB \cite{liu2007bindingdb}, and one ABFE score of a compound randomly sampled from the same BindingDB dataset. The initial dataset for c-MET is the same, except the experimental data contains 102 experimental activity values from BindingDB's c-MET target.

We initialize our dataset with only one datapoint from ABFE so that we can make maximal use of active learning. The candidate set, which is the pool of compounds available to active learning, is obtained from BindingDB under the ``BRD4'' target for BRD4(2) and ``c-MET'' for c-MET. Note that the BRD4 target is a larger superset of BRD4(2), which we chose because there were not enough compounds measured against only BRD4(2).

Our held-out test set consists of precomputed ABFE results for a curated set of 32 compounds for each target. Half of these compounds come from BindingDB's BRD4(2) or c-MET dataset, with the constraint that the experimental activity be $<1 \mu M$, and the other half from BRD4(2) or c-MET decoys, which are presumed to be inactive, generated by DUD-E \cite{mysinger2012directory}. To ensure diversity, we clustered BindingDB compounds in the test set so that no compound had a Tanimoto similarity higher than $0.4$ with any other compound, and then generated decoy counterparts for each of these compounds using DUD-E. We also ensured the test compounds were dissimilar (Tanimoto similarity $<0.4$) to the compounds in the initial training and candidate sets, removing them from the latter if any compounds were too similar. We constructed the test dataset with both actives and decoys so that we could measure the ability of each simulator to distinguish between them (see Appendix \ref{appendix-result-environment}).


\subsubsection{Baselines}
We compare our model with the following baseline methods for multi-fidelity surrogate modeling:

\begin{itemize}
    \item \textbf{Only ABFE (NN).} A simple feedforward neural network that is only trained on ABFE scores.
    \item \textbf{Direct-GP (DKL).} A DKL-based \cite{wilson2016deep} multi-fidelity GP model using a downsampling kernel for the fidelities \cite{wu2020practical}. Uses the ``Direct'' output from AutoDock4, meaning the total binding energy prediction, and discards all other AutoDock4 outputs.
    \item \textbf{Surrogate-GP (DKL).} Same as above, except for $f_2$ we use a linear ``Surrogate'' model that takes all outputs from AutoDock4 as input and outputs a prediction of the ABFE score.
    \item \textbf{D-MFDAL} \cite{wu2023disentangled}. A state-of-the-art neural process model and acquisition function for multi-fidelity modeling. Learns a global and local representation for each fidelity level, avoiding the propagation of errors from lower to higher fidelity levels.
    \item \textbf{DMFAL} \cite{li2020deep}. A neural network-based approach for modeling multi-fidelity high-dimensional outputs by passing information from lower to higher fidelity levels.
    \item \textbf{Hadamard-MT (DKL)} \cite{bonilla2007multi}. A DKL-based multi-task (MT) GP model where the kernel is the Hadamard product of an input and task kernel. Each fidelity is treated as its own task, except for AutoDock4 where we use all 16 outputs by treating each one as its own task.
\end{itemize}

See Appendix \ref{appendix-baselines} for further details about these baselines.

\subsubsection{Results}
\begin{figure*}[t!]
\begin{center}
\centerline{\includegraphics[width=\textwidth]{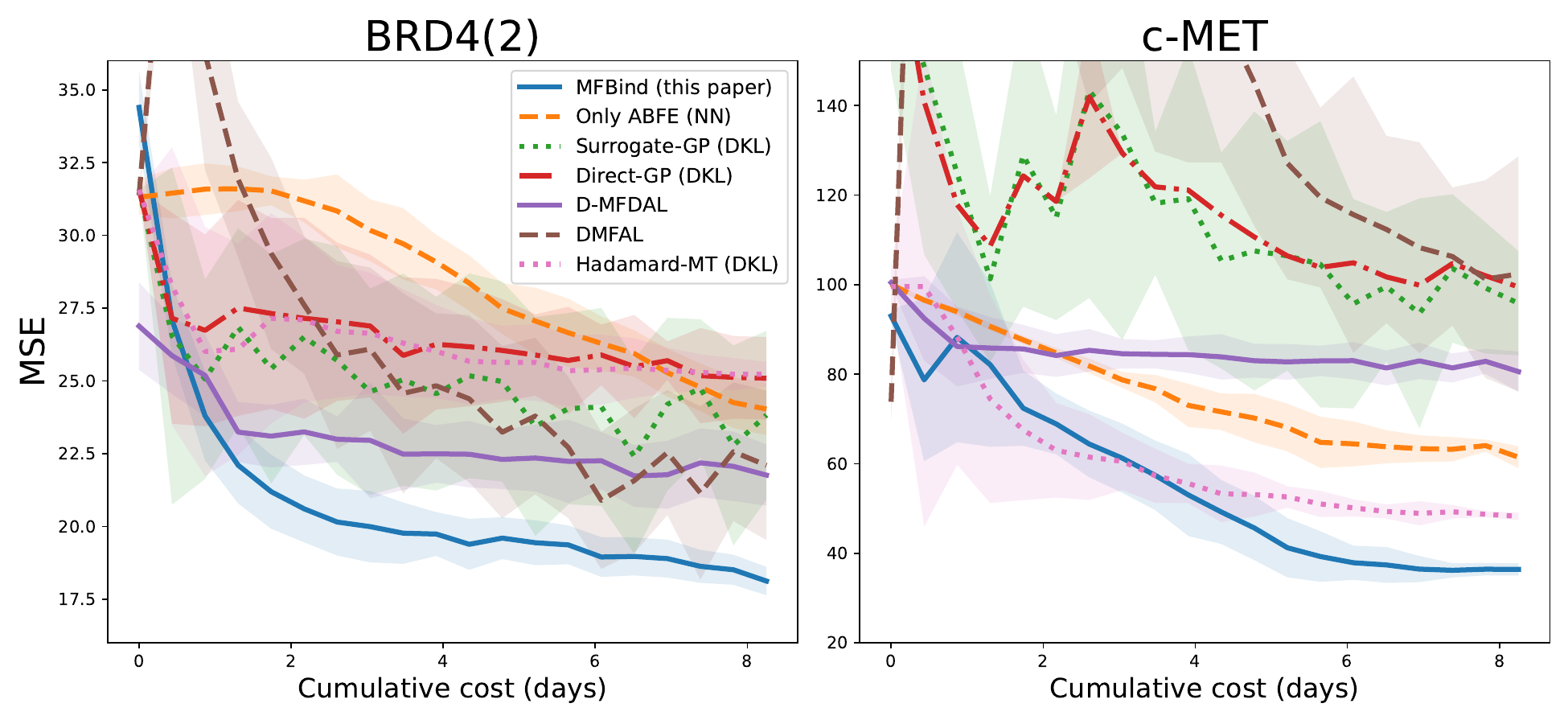}}
\vskip -0.1in
\caption{\textbf{Regression of ABFE scores in an active learning setting}. The y-axis shows the mean squared error (MSE), in kcal/mol, of each method on the held-out test set. The x-axis shows the cumulative active learning query cost in days (wall clock time on a 9 core, 8 GPU server). Each line represents an average over 20 runs with random seeds (using caching of ABFE results to reduce running times), with the shaded region indicating the standard deviation across runs.}
\label{fig3}
\end{center}
\vskip -0.2in
\end{figure*}

\paragraph{Surrogate model performance.} Figure \ref{fig3} shows the prediction error for both targets, measured in MSE between the actual and predicted ABFE results in kcal/mol for the 32 compounds in the held-out test set, as the computational budget allotted to active learning increases. The single-fidelity, only ABFE approach performs poorly for both targets, showing that MFBind aids in training models to predict ABFE scores at a lower computational cost than using only ABFE data. Among the multi-fidelity surrogate modeling baselines, ours performs the best across both targets, suggesting that it is the most efficient at using cheaper non-ABFE methods to enhance the prediction of ABFE results. 

\paragraph{Ablation study.} Table \ref{ablation-table} (first set of rows) shows the performance of various ablations of the surrogate model, measured by MSE in kcal/mol on the test set. Each method was trained on the same data as collected by MFBind during active learning. ``w/o Pretraining'' means we did not perform any pretraining on the lower fidelity data, and instead trained the model on the full number of epochs using all fidelities. ``w/o Regularization'' means we removed the linear prediction head regularization term. These results show that both design choices, especially the pretraining step, significantly contribute to the performance of our model.

We also studied the impact of ablating each lower fidelity simulator (Table \ref{ablation-table}, second set of rows). ``w/o AutoDock4 data'' means we removed all data from the AutoDock4 fidelity level, and ``w/o Experimental data'' means we removed all experimental data. As demonstrated, both of the lower fidelity simulators are important for the model to learn to predict data at the highest fidelity level.

\begin{table}
\vskip -0.1in
\caption{\textbf{MFBind ablations}. We report the test set MSE, for each target, of each model modification when trained on the same set of data queried by the full MFBind model at the end of the active learning experiment.}
\vskip 0.1in
\label{ablation-table}
\begin{center}
\begin{small}
\begin{sc}
\begin{tabular}{l|cc}
\toprule
Ablation & BRD4(2) & c-MET\\
\midrule
None (MFBind) & \textbf{18.1} & \textbf{36.0}\\
w/o Pretraining & 23.7 & 41.8\\
w/o Regularization & 18.8 & 38.8\\
\midrule
w/o AutoDock4 data & 20.4 & 37.5\\
w/o Experimental data & 25.5 & 40.8\\
\bottomrule
\end{tabular}
\end{sc}
\end{small}
\end{center}
\end{table}

\subsection{Compound generation with MFBind}

\label{exp-generative}

\subsubsection{Setup}

To show the utility of MFBind in assisting molecular generative modeling, we seek to generate compounds using the MFBind surrogate model as the reward function. We chose LIMO \cite{eckmann2022limo}, a variational autoencoder-based generative model, due to its strong performance on binding affinity optimization. We also explored MolDQN \cite{zhou2019optimization}, a reinforcement learning approach, as the generative model. However, we found MolDQN generated compounds that were not drug-like and in some cases chemically implausible, even when including a drug-likeness objective in the reward (see Appendix \ref{appendix-moldqn-section}). 

We first trained ours and baseline surrogate models on the same initial training dataset as the previous active learning task, except now including the ABFE scores for all compounds in the candidate set. This ensured that the predictions from the reward function were as accurate as possible. Then, we froze the surrogate model and used it as a reward function to evaluate generated compounds. We experimented with periodically updating the surrogate model with active learning over the generated compounds, but found that it led to comparable or worse performance than not updating the model.


For each target and choice of surrogate model, we used LIMO to generate 10,000 compounds intended to bind the target, chose the top 20 unique compounds with the highest surrogate-computed reward, and used full ABFE calculations to estimate their binding affinities. We incorporated an additional drug-likeness (QED; \citet{bickerton2012quantifying}) objective in the reward function, by adding 2x the QED of the compound to the ABFE score predicted by the surrogate model, to ensure the generated compounds were reasonably drug-like. We also enforced a QED cutoff $>0.5$ and maximum ring size $<7$ for choosing the final 20 compounds.

\subsubsection{Baselines}
We compared compounds generated with MFBind as the reward function against the following baseline reward functions:
\begin{itemize}
    \item \textbf{Single fidelity (SF) ABFE}. A single-fidelity surrogate model that is trained only on ABFE data. This baseline is trained on the same number of ABFE datapoints as MFBind, but without the data from the other fidelities.
    \item \textbf{Single fidelity (SF) AutoDock4}. Same as above, except this baseline is only trained on the AutoDock4 total binding energy score without any other fidelity data, including ABFE. This baseline is the prevalent approach in molecular generative modeling, where the docking score is used as a reward function.
\end{itemize}
See Appendix \ref{appendix-baselines} for more details about each baseline. We did not include other multi-fidelity surrogate modeling baselines in this experiment because of computational constraints, and the high noise in the generative setting which makes a comparison between competing multi-fidelity surrogates difficult without a large number of samples.

\subsubsection{Results}

\begin{table*}
\vskip -0.1in
\caption{\textbf{Evaluation of LIMO-generated compounds with different surrogate models.} The mean and top 3 ABFE-computed energies, both in kcal/mol, are shown among 20 tested compounds from each method and for each target. ``SF'' refers to single-fidelity methods that only use one simulator, while our ``MFBind'' approach uses all simulators. All compounds have QED $>0.5$.}
\label{gen-table}
\vskip 0.11in
\begin{center}
\begin{small}
\begin{sc}
\begin{tabular}{l|cccc|cccc}
\toprule
& \multicolumn{4}{c}{BRD4(2)} & \multicolumn{4}{c}{c-MET}\\
\midrule
Method & Mean & 1st & 2nd & 3rd & Mean & 1st & 2nd & 3rd\\
\midrule
SF ABFE & -2.94 & -5.60 & -5.26 & -4.57 & 3.14 & -5.19 & -3.05 & -2.79\\
SF AutoDock4 & -2.81 & -4.46 & -4.01 & -3.40 & 2.81 & -5.87 & -3.30 & -2.51\\
MFBind & \textbf{-4.16} & \textbf{-10.94} & \textbf{-10.04} & \textbf{-7.38} & \textbf{-3.69} & \textbf{-11.25} & \textbf{-11.06} & \textbf{-9.10}\\
\bottomrule
\end{tabular}
\end{sc}
\end{small}
\end{center}
\vskip -0.1in
\end{table*}

Table \ref{gen-table} shows the computed ABFE scores for the top 20 compounds with the highest predicted reward generated under the guidance of each surrogate model for each target. For both targets, among the sample of 20 compounds, the mean ABFE score of compounds from MFBind is distinctly lower (indicating better binding affinity) than the same number of compounds generated using either the single-fidelity ABFE or single-fidelity AutoDock4 approaches. Additionally, the top three compounds for both targets from MFBind have dramatically better ABFE results than the top three compounds from the single-fidelity methods. Note that MFBind is the only approach that generated compounds in the nanomolar $K_d$ range ($<-8.2$ kcal/mol), a widely used activity cutoff in early drug discovery to determine which compounds show promise \cite{hughes2011principles}. 

\begin{figure}
\vskip 0.1in
\begin{center}
\centerline{\includegraphics[width=0.9\columnwidth]{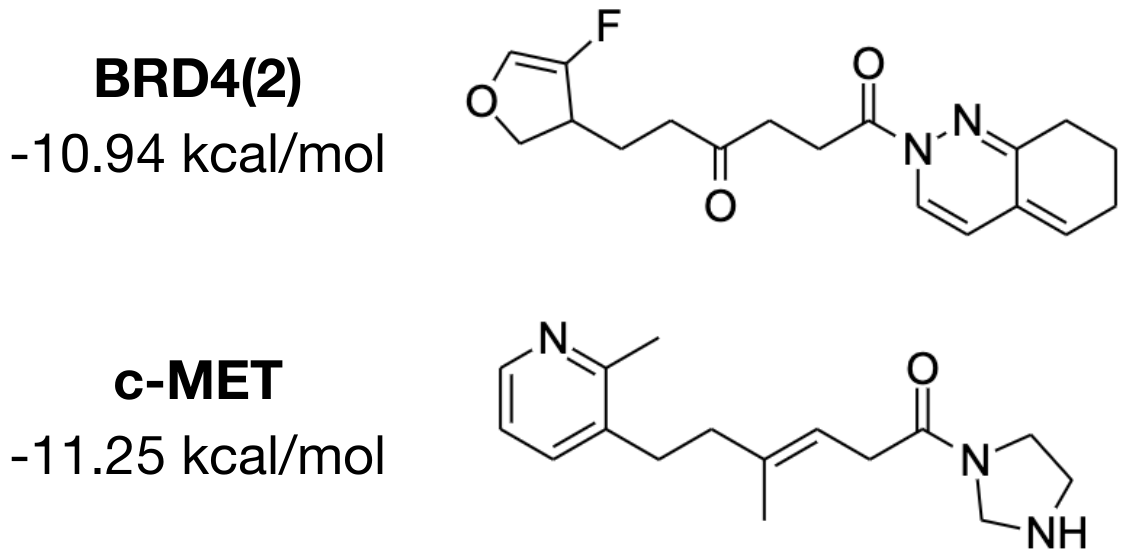}}
\caption{\textbf{Selected generated compounds from LIMO + MFBind}. The top compound for both BRD4(2) and c-MET are shown. See Appendix \ref{appendix-limo-compounds-sec} for more compounds.}
\label{main-text-limo-compounds-fig}
\end{center}
\vskip -0.2in
\end{figure}

Figure \ref{main-text-limo-compounds-fig} shows the top compound generated from LIMO + MFBind for each target. The compounds appear relatively synthnesizable and drug-like while having strong ABFE-predicted affinity. See Appendix \ref{appendix-limo-compounds-sec} for more examples.

These results, combined with the fact that ABFE results correlate well with experiment for both targets \cite{heinzelmann2021automation, huggins2022comparing}, indicate that the MFBind surrogate model allows us to generate much more promising compounds than either of the single-fidelity approaches.
\section{Discussion and Conclusion}
We present a new multi-fidelity framework, MFBind, for evaluating compound binding affinity in generative models. We introduce a new multi-fidelity environment for binding affinity that consists of docking (AutoDock4), experimental data (from BindingDB), and binding free energy (ABFE) simulators. Our framework also contains a deep surrogate model trained with an active learning algorithm to further reduce cost. Our surrogate model can fit small amounts of high-fidelity data using a pretraining method on the lower fidelity data, combined with regularized linear fidelity-specific prediction heads. 

We perform extensive evaluation of ours and baseline approaches in multi-fidelity surrogate modeling, and find that our model is most capable of efficiently utilizing a limited computational budget to predict the ABFE score of unseen compounds. We also test our framework in a molecular generative modeling task, where we use the surrogate model as a reward function for generation. We find that MFBind outperforms common approaches that use a single-fidelity reward function.  It can generate compounds with markedly higher activity, as computed by the accurate binding free energy simulator, than competing methods. Therefore, MFBind shows promise as a way to make generative models for drug discovery useful in practice.

We note that the greatest accuracy surrogate model would presumably be obtained by training purely against, e.g., thousands of ABFE results, without the other fidelities. However, generating these training data would be exceedingly costly, given that ABFE calculations are around 1250x slower than docking calculations. An important implication of the present results is that, although docking results are not as accurate as ABFE results, and although their correlation coefficients with ABFE results are modest ($r=0.25$ (BRD4(2)) and $r=0.23$ (c-MET), see Appendix \ref{appendix-additional-exp}), the docking results can still be used to boost the predictivity of a model trained with a fixed number of ABFE results.

Limitations of our approach include a limited set of simulators and potentially a lack of synthesizability of the generated molecules. Since the only objectives we consider when generating compounds are the binding affinity and QED, it is possible that the compounds could be difficult to synthesize. Additionally, our acquisition function for active learning is somewhat simple. Instead of averaging the uncertainty from all AutoDock4 outputs, further studies can be done on weighing them by importance.

Future work could include making the acquisition function more complex, and adding more fidelities, such as deep learning-based binding affinity predictors and ABFE with varying simulation times. Another next step would be to use a reaction-aware generative model that generates more synthesizable molecules, such as \citet{horwood2020molecular}.

\paragraph{Impact statement}
Similar to other works that apply machine learning to drug discovery, our work is subject to dual use \cite{urbina2022dual}. There is potential for societal benefit, by helping develop new drug compounds to treat disease. However, there is also potential for harm, such as to generate new chemical weapons. Fortunately, the latter places a whole additional set of requirements on compounds (e.g. skin-absorbable or volatile and subject to inhalation), so this problematic direction does not appear to be imminent. 

\section{Acknowledgments}
This work was supported in part by Army-ECASE award W911NF-07-R-0003-03, the U.S. Department Of Energy, Office of Science, IARPA HAYSTAC Program, CDC-RFA-FT-23-0069, NSF Grants \#2205093, \#2146343, and \#2134274. MKG has an equity interest in and is a cofounder and scientific advisor of VeraChem LLC.

\bibliography{main.bib}
\bibliographystyle{icml2024}

\newpage
\appendix
\onecolumn
\section{Environment details}
\label{appendix-env}

For all simulators, we estimated the cost using the average over 10 samples with random input compounds.

\paragraph{AutoDock4} We prepared the AutoDock4 grid files using AutoDockTools \cite{morris2009autodock4}. Arbitrary ligands were prepared using obabel \cite{o2011open} with pH 7.4 and gasteiger partial charges. We used AutoDock-GPU \cite{santos2021accelerating}, a GPU-accelerated version of AutoDock4, for all computation. The full set of outputs we collected from AutoDock4 are as follows, with the last 9 collected in a post-processing step from BINANA \cite{young2022binana}:
\begin{itemize}
    \item Total binding energy (``Estimated Free Energy of Binding'' in the AutoDock4 output), minimum over 20 random restarts
    \item Total binding energy, mean over 20 random restarts
    \item Intermolecular energy
    \item Internal energy
    \item Torsional energy
    \item Unbound system energy
    \item Number of ligand atoms
    \item Number of protein-ligand hydrogen bonds
    \item Number of protein-ligand pi-pi stacking bonds
    \item Number of protein-ligand salt bridges
    \item Number of protein-ligand T-stacking interactions
    \item Number of protein-ligand close contacts
    \item Backbone alpha flexibility
    \item Backbone other flexibility
    \item Sidechain alpha flexibility
    \item Sidechain other flexibility
\end{itemize}

\paragraph{Absolute binding free energy (ABFE)} We use the Binding Affinity Tool (BAT.py) implementation \cite{heinzelmann2021automation} for absolute binding free energy calculation, available at \url{https://github.com/GHeinzelmann/BAT.py}, which uses the simultaneous decoupling and recoupling (SDR) method. All molecular dynamics simulators are run with AMBER with GPU support. As BAT.py requires a starting pose for the ligand, we used the pose generated from AutoDock4. We found that we were able to reduce the simulation times up to 80\% from the default times for each phase of the SDR computation without losing much accuracy. We additionally wrote custom scripts to parallelize molecular dynamics runs across all available GPUs.  

\section{Experimental details}
\label{appendix-experimental-details}

All experiments were conducted on a server with 8 RTX 2080 Ti GPUs. For our model and each baseline, we performed a random hyperparameter search with 20 trials (across the hyperparameters listed below) and took the combination with the best test set MSE when trained on the initial dataset combined with ABFE datapoints for the entire candidate set. We conducted separate hyperparameter searches for each target, and used the same set of hyperparameters for both the surrogate modeling and generative experiments. All models were trained with the Adam optimizer.

\subsection{MFBind surrogate model details}

\label{appendix-mfbind}

\begin{figure*}[ht]
\begin{center}
\centerline{\includegraphics[width=0.4\textwidth]{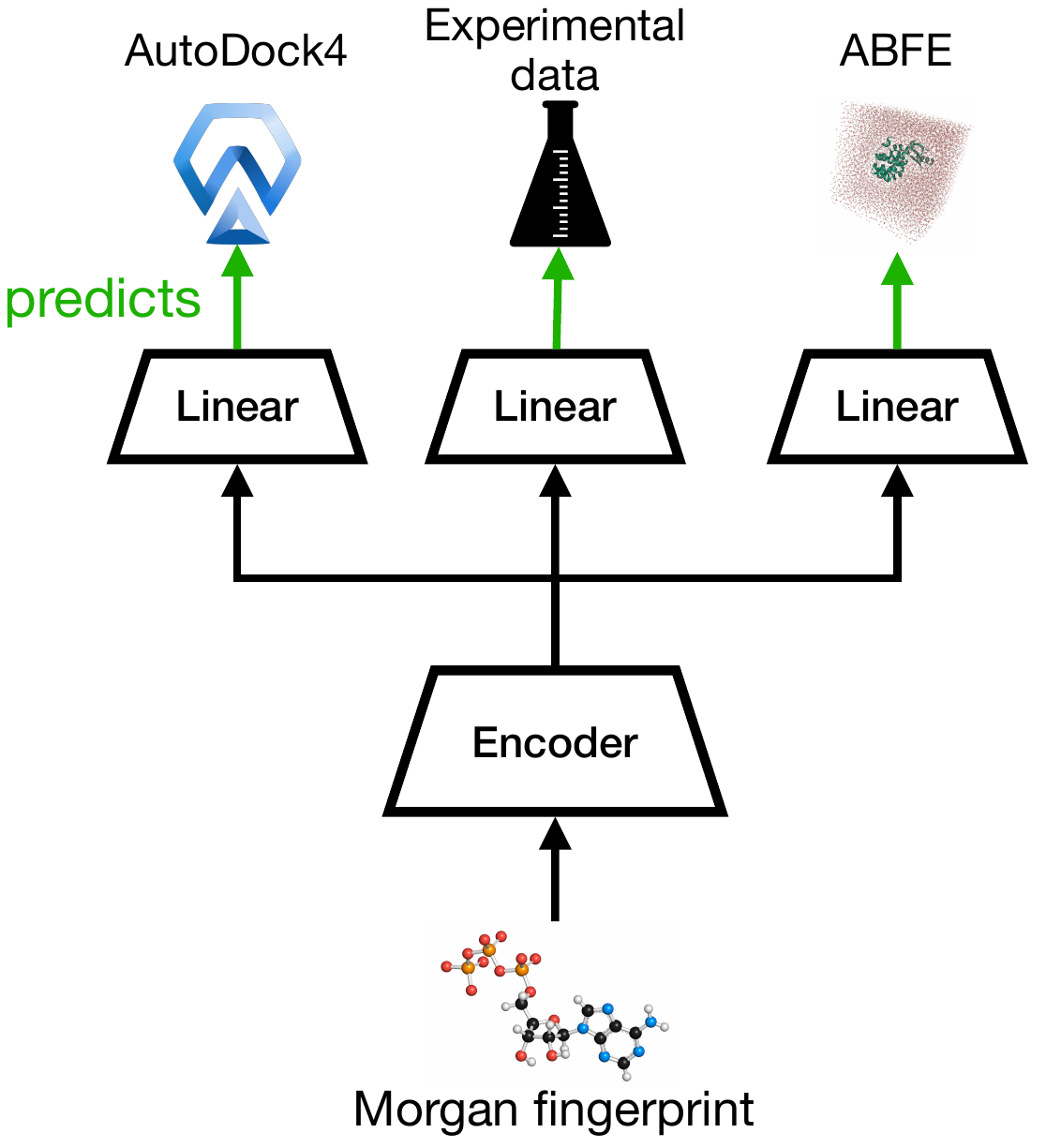}}
\caption{\textbf{Diagram of the MFBind surrogate model.} An input molecule, represented as a Morgan fingerprint, is fed through the deep encoder to produce a latent representation. That representation is then passed to linear fidelity-specific prediction heads to produce a prediction for each fidelity level.}
\end{center}
\label{appendix-mfbind-surrogate-fig}
\end{figure*}

Figure \ref{appendix-mfbind-surrogate-fig} is a diagram of the MFBind surrogate model. Our deep molecular encoder consisted of 4 linear layers with ReLU activations and dropout after each layer, except the final layer. For estimating the model uncertainty, we used the predictive variance over 50 samples using the same input with random dropout. Hyperparameters: \texttt{encoding dim($n$), hidden layer dim, lr, lr decay beta, $\lambda_0$, $\lambda_1$, $\lambda_2=1$, $\lambda_{reg}$, pretraining epochs, finetuning epochs, dropout p}

\subsection{Baseline details}
\label{appendix-baselines}

\subsubsection{Multi-fidelity surrogate modeling}

\paragraph{Only ABFE (NN)} Simple feedforward neural network using the same architecture of molecular encoder as MFBind, and a final linear layer to produce the ABFE prediction. Only trained on ABFE data. Used the same MC dropout technique as our model to estimate the uncertainty. Hyperparameters: \texttt{encoding dim, hidden layer dim, lr, lr decay beta, num epochs, dropout p}

\paragraph{Direct-GP (DKL)} Exact GP model using a 3-layer deep kernel with ReLU activations to encode the input molecule, which is then passed to the GP. A downsampling kernel \cite{wu2020practical} is used to produce the output at each fidelity level. Since this model can only fit a scalar for each fidelity level, we used the total energy prediction from AutoDock4 only (instead of all 16 outputs). For this and all other GP-based baselines, we used the posterior variance to estimate model uncertainty. Implemented using the BoTorch \cite{balandat2020botorch} library. Hyperparameters: \texttt{encoding dim, DKL hidden layer dim, lr, num epochs}

\paragraph{Surrogate-GP (DKL)} Same as above, except instead of using the total energy prediction from AutoDock4, we used a linear surrogate. Specifically, we took the available ABFE datapoints and trained a linear regression model with L2 regularization to predict the ABFE score of a compound using all outputs from AutoDock4 as input. Then, this model was applied to all AutoDock4 datapoints to transform the multi-dimensional output into a scalar for use in the GP model. Hyperparameters: same as above

\paragraph{D-MFDAL} Defines its own acquisition function, which we used. Used the code available at \url{https://github.com/Rose-STL-Lab/Multi-Fidelity-Deep-Active-Learning}. Hyperparameters: \texttt{hidden dim, epoch num, lr}

\paragraph{DMFAL} Defines its own acquisition function, which we used. Used the code available at \url{https://github.com/shib0li/DMFAL}. Hyperparameters: \texttt{lr, reg strength, max epoch, hidden layer dim, base dim}

\paragraph{Hadamard-MT (DKL)} Exact GP model using a 3-layer deep kernel with ReLU activations to encode the input molecule, which is then passed to the GP. This multi-task GP uses the Hadamard product of the input kernel and a task kernel \cite{bonilla2007multi}. The task index is concatenated to the input (from the deep kernel). Since this is a multi-task model, we can model all outputs from AutoDock4 by treating each one as its own task. Therefore, we learned 18 total tasks: 1 for experimental data, 1 for ABFE, and 16 for AutoDock4. Implemented using GPyTorch \cite{gardner2018gpytorch}. Hyperparameters: \texttt{encoding dim, DKL hidden layer dim, lr, num epochs}

\subsubsection{Compound generation with MFBind}
For these baselines, we used the same hyperparameters as those chosen for the Only ABFE (NN) baseline in the above surrogate modeling task.

\paragraph{Single fidelity (SF) ABFE} Simple feedforward neural network using the same architecture of molecular encoder as MFBind, and a final linear layer to produce the prediction. Only trained on ABFE data.

\paragraph{Single fidelity (SF) AutoDock4} Same as above, except only trained on the total binding energy prediction from AutoDock4 (without the 15 other outputs).

\section{Additional results}
\label{appendix-additional-exp}

\subsection{Analysis of MFBind environment}
\label{appendix-result-environment}

\begin{figure*}[h]
\begin{center}
\centerline{\includegraphics[width=\textwidth]{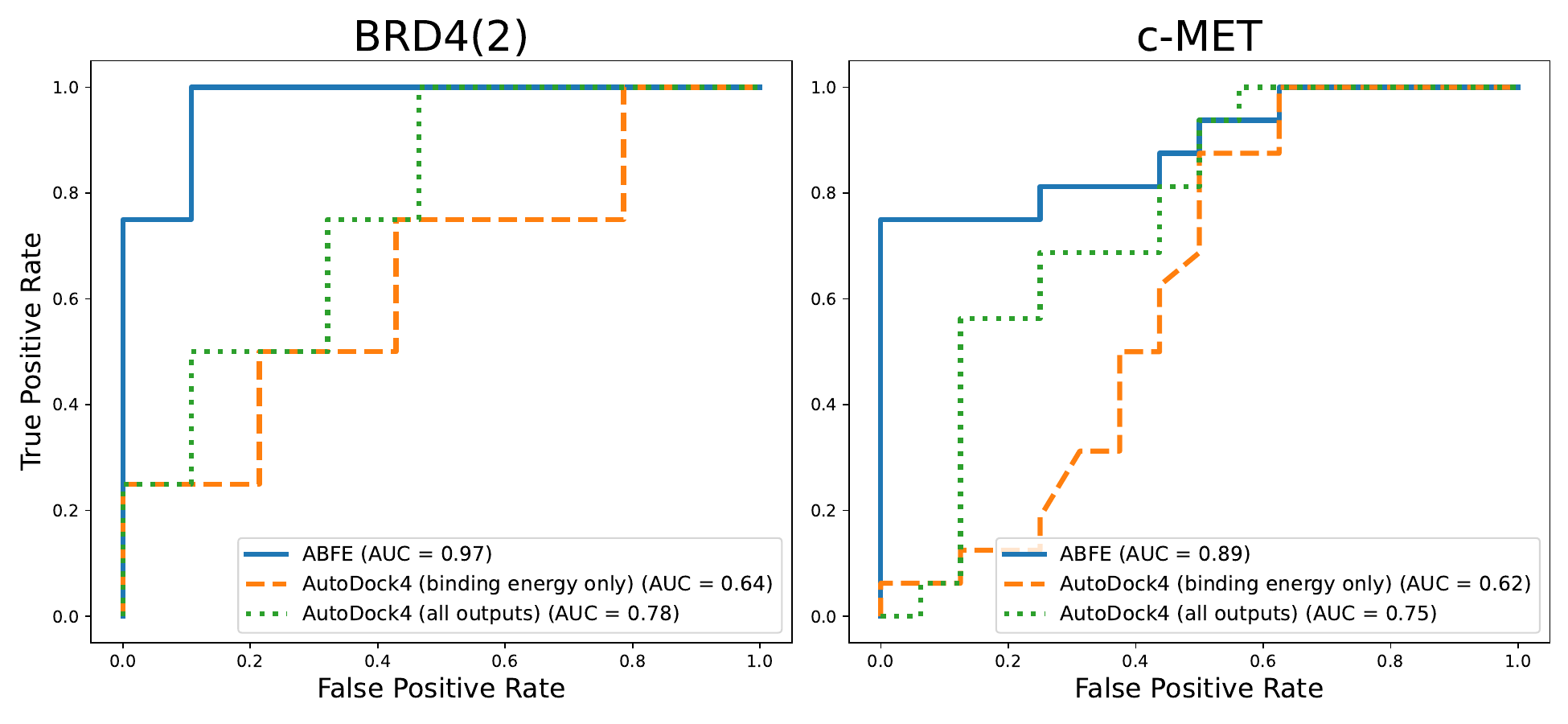}}
\caption{\textbf{ROC curve of each simulator.} Compounds from the BindingDB BRD4(2) and c-MET datasets are classified as active, and decoys are classified as inactive.}
\label{fig2}
\end{center}
\end{figure*}

We aim to show that each simulator in the MFBind environment meaningfully correlates to real-world experimental data. Figure \ref{fig2} shows a ROC curve for the ABFE and AutoDock4 simulators on our BRD4(2) and c-MET test datasets. The curve for experimental data is not shown, even though it is its own ``simulator'', because there is no experimental data for half of the test set (the decoy compounds). We classified all compounds from the BindingDB BRD4(2) and c-MET datasets, which had activity $<1 \mu M$, as ``Active'', and all other decoy compounds as ``Inactive.'' ``AutoDock4 (binding energy only)'' uses the total binding energy prediction from AutoDock4, while ``AutoDock4 (all outputs)'' uses a linear surrogate model that takes all outputs from AutoDock4 as input and outputs a prediction of the ABFE score.

As shown, ABFE is the most predictive data source for experimental data, with a ROC-AUC of 0.97 for BRD4(2) and 0.89 for c-MET. As expected, the total binding energy from AutoDock4, the computationally cheaper data source, is a worse predictor, but is still moderately predictive (ROC-AUC of 0.64 and 0.62). AutoDock4 (all outputs), which uses all outputs from AutoDock4 to make predictions, is more predictive (ROC-AUC of 0.78 and 0.75) than AutoDock4 total binding energy, but still worse than ABFE. We also measured the correlation between the binding energy predictions from AutoDock4 and ABFE, finding a correlation of $r=0.25$ for BRD4(2) and $r=0.23$ for c-MET. This helps explain why the multi-fidelity approach works, because the cheaper simulator is correlated with the more expensive simulator.

These results show that the MFBind environment has the desirable property that the more expensive simulators make more accurate predictions, meaning that it holds promise as an approach to making high-quality ABFE predictions without incurring an infeasibly high computational cost. They also help motivate the multi-output approach with AutoDock4, because it appears that considering all outputs from AutoDock4 is more useful than just the total energy prediction.

\subsection{Compounds from LIMO generation}
\label{appendix-limo-compounds-sec}

\begin{figure}[H]
\begin{center}
\centerline{\includegraphics[width=\columnwidth]{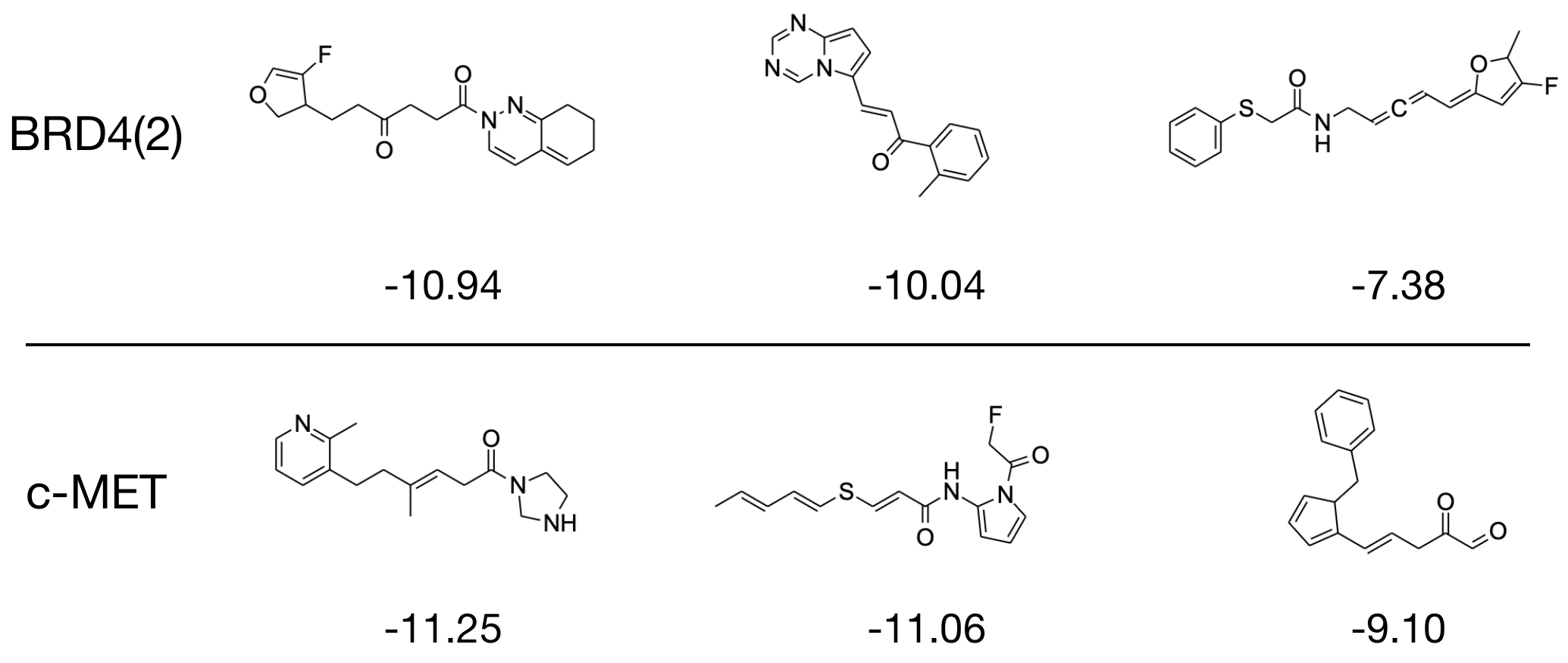}}
\caption{\textbf{Generated compounds from LIMO + MFBind}. The ABFE score of each compound is shown at the bottom. The top row shows the top 3 compounds generated for BRD4(2), and the bottom row for c-MET.}
\label{appendix-limo-compounds-fig}
\end{center}
\end{figure}

Figure \ref{appendix-limo-compounds-fig} shows the top 3 compounds for each target generated from the LIMO generative model \cite{eckmann2022limo} with MFBind as the reward function. As shown, the generated compounds are relatively drug-like while showing favorable binding.

\subsection{Generation with MolDQN}

\label{appendix-moldqn-section}

Table \ref{appendix-moldqn-table} shows the results from the same experimental procedure as from Section \ref{exp-generative}, except using MolDQN as the generative model. We only tested the BRD4(2) target. While we decided not to include these results in the body text due to the generated compounds not being very drug-like, these results show that MFBind is superior to the single-fidelity approaches on another generative model. The lack of drug-likeness is likely due to the nature of the generative model, and cannot be attributed to the surrogate model used as the reward.

\begin{figure}[H]
\begin{center}
\centerline{\includegraphics[width=\columnwidth]{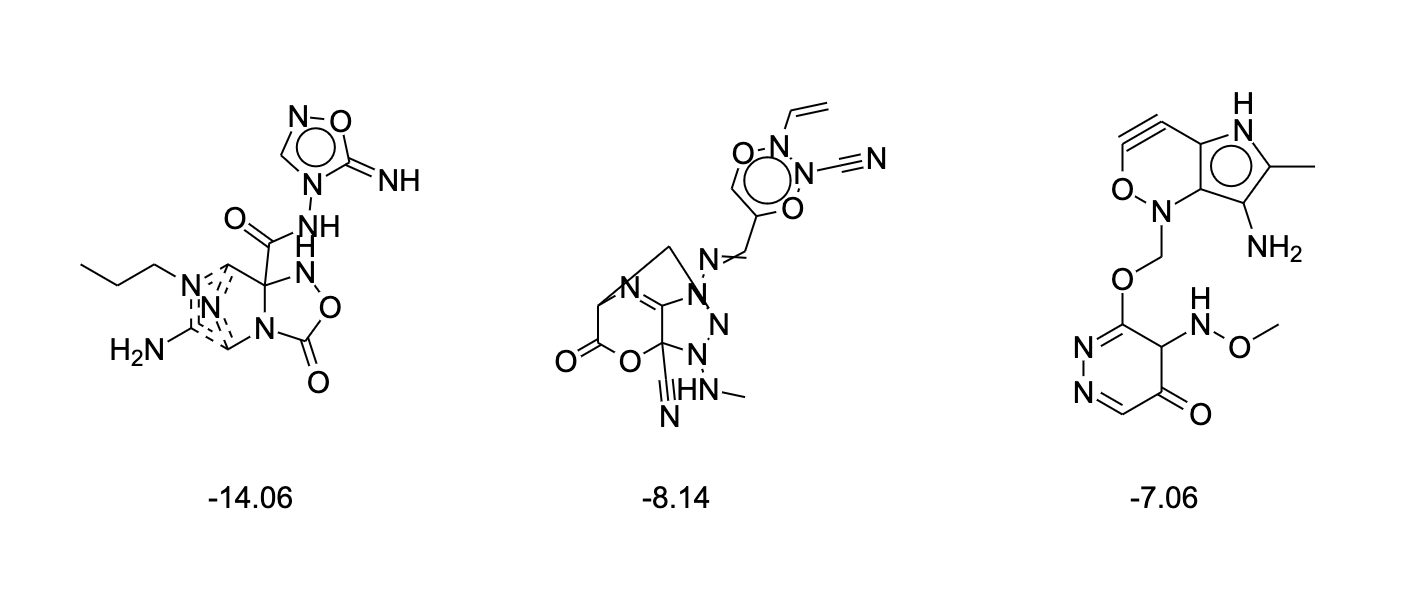}}
\caption{\textbf{Generated compounds from MolDQN + MFBind}. The ABFE score of each compound is shown at the bottom. The compounds are generally non drug-like, and some are chemically implausible (e.g. the rightmost compound with a triple bond in a ring).}
\label{appendix-moldqn-compounds-fig}
\end{center}
\end{figure}

Figure \ref{appendix-moldqn-compounds-fig} shows the top 3 generated compounds from MolDQN with MFBind as the reward function. The compounds are generally non drug-like and have implausible structures such as the triple bond in a ring on the rightmost compound.

\begin{table}[H]
\caption{\textbf{Evaluation of MolDQN-generated compounds for BRD4(2).} The mean and top 3 ABFE-computed energies are shown among 20 tested compounds from each method. ``SF'' refers to single-fidelity methods that only use one simulator, while our ``MFBind'' approach uses all simulators. All compounds have QED $>0.5$.}
\label{appendix-moldqn-table}
\begin{center}
\begin{small}
\begin{sc}
\begin{tabular}{l|cccc}
\toprule
Method & Mean & 1st & 2nd & 3rd\\
\midrule
SF ABFE & 1.78 & -6.52 & -4.47 & -3.89\\
SF AutoDock4 & -1.14 & -6.66 & -6.64 & -5.41\\
MFBind & \textbf{-3.41} & \textbf{-14.06} & \textbf{-8.14} & \textbf{-7.06}\\
\bottomrule
\end{tabular}
\end{sc}
\end{small}
\end{center}
\end{table}

\end{document}